\documentclass{article}
\usepackage{spconf,amsmath,graphicx,hyperref}

\usepackage{algorithm}
\usepackage{algpseudocode}
\usepackage{subcaption}
\usepackage{float}
\usepackage{stfloats}
\usepackage{enumitem}
\usepackage{cancel}
\usepackage{amsfonts}

\let\OLDthebibliography\thebibliography
\renewcommand\thebibliography[1]{
  \OLDthebibliography{#1}
  \setlength{\parskip}{0pt}
  \setlength{\itemsep}{0.4em}
}

\title{Domino: Dominant Path-based Compensation for Hardware Impairments in Modern WiFi Sensing}
\name{Ruiqi Kong and He Chen}
\address{Department of Information Engineering, The Chinese University of Hong Kong, Hong Kong SAR, China. \\E-mail: \{rqkong, he.chen\}@ie.cuhk.edu.hk}

\begin{document}
%
\maketitle
\begin{abstract}
WiFi sensing faces a critical reliability challenge due to hardware-induced RF distortions, especially with modern, market-dominant WiFi cards supporting 802.11ac/ax protocols. These cards employ sensitive automatic gain control and separate RF chains, introducing complex and dynamic distortions that render existing compensation methods ineffective. In this paper, we introduce Domino, a new framework that transforms channel state information (CSI) into channel impulse response (CIR) and leverages it for precise distortion compensation. Domino is built on the key insight that hardware-induced distortions impact all signal paths uniformly, allowing the dominant static path to serve as a reliable reference for effective compensation through delay-domain processing. Real-world respiration monitoring experiments show that Domino achieves at least $2\times$ higher mean accuracy over existing methods, maintaining robust performance with a median error below 0.24 bpm, even using a single antenna in both direct line-of-sight and obstructed scenarios.

\end{abstract}
\section{Introduction}
\label{sec:intro}

WiFi sensing leverages existing WiFi infrastructure to extract subtle environmental changes from radio frequency (RF) signals, offering advantages over traditional sensing approaches that require dedicated sensors or wearable devices \cite{intell22zeng}. 
The availability of fine-grained channel state information (CSI) through commercial off-the-shelf (COTS) devices has democratized WiFi sensing research \cite{wifiact24lyons}. 
Recent advances in CSI-based WiFi sensing have demonstrated promising applications in healthcare monitoring \cite{wang23improved,multi24chang}, human activity recognition \cite{meneghello2022sharp,joint23saleh}, and indoor localization \cite{kotaru2015spotfi,qian2018widar,zhang2022nlos}, attracting significant attention from both academia and industry.

\begin{figure}
    \centering
    
    \includegraphics[width=\linewidth]{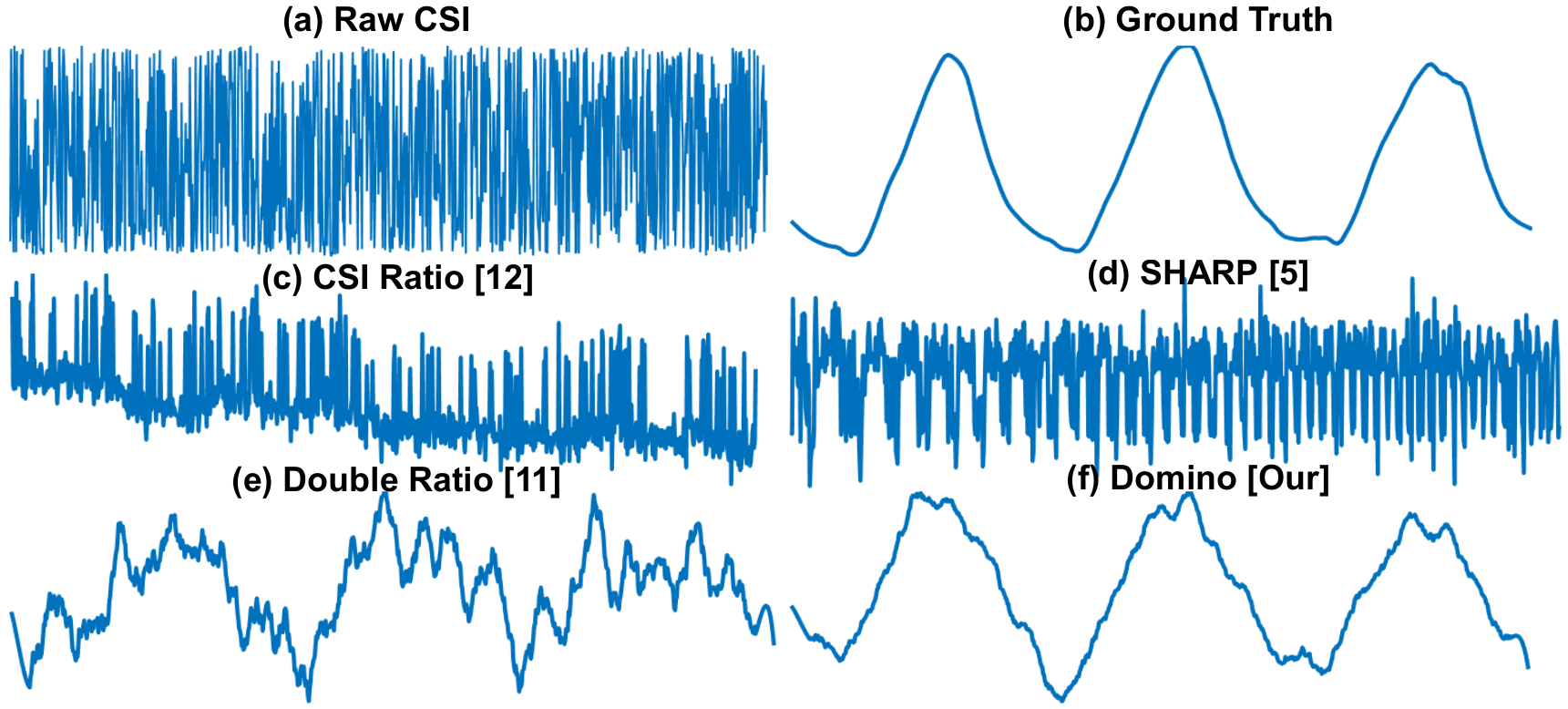}
    \caption{Comparison of studies on distortion compensation.}
    \label{fig:dc_comp}
    \vspace{-1.5em}
\end{figure}

Despite significant progress, WiFi sensing faces a critical transition challenge that threatens its reliability and practical viability. Reliable WiFi sensing critically depends on high-fidelity sensing signals, which are largely affected by RF distortions in COTS devices. Hardware impairments introduce time-varying distortions in both magnitude and phase of CSI measurements \cite{terry2002ofdm}. These distortions create significant challenges by obscuring the subtle CSI variations induced by target activities. This challenge has become particularly acute with the market dominance of modern WiFi cards supporting 802.11ac/ax protocols \cite{enabling24yi}. While these cards offer enhanced performance through wider bandwidth and advanced techniques like multi-user MIMO, they employ sensitive automatic gain control (AGC) and separate RF chains for multiple antennas, introducing more complex distortions that render existing compensation methods ineffective. 

We illustrate this phenomenon in Fig.~\ref{fig:dc_comp} using CSI measurements collected during human respiration scenarios. Fig.\ref{fig:dc_comp}a displays the magnitude variation of raw CSI collected using a modern WiFi card (Intel AX200) during human respiration, with Fig.~\ref{fig:dc_comp}b showing the corresponding ground-truth breathing pattern. The CSI magnitude variations are clearly corrupted by severe time-varying distortions. Fig.~\ref{fig:dc_comp}c-e show the compensated temporal CSI patterns of the subcarrier exhibiting the highest periodicity, as processed by various existing methods. Specifically, CSI ratio-based techniques \cite{zeng2019farsense, indoor17li} fail to address antenna-specific distortions introduced by separate RF chains  (Fig.~\ref{fig:dc_comp}c). SHARP's optimization-based method~\cite{meneghello2022sharp} cannot accurately estimate multipath components,  insufficient for fine-grained respiration sensing (Fig.~\ref{fig:dc_comp}d). Although the reference subcarrier (i.e., double ratio) approach \cite{enabling24yi} shows improvement (Fig.~\ref{fig:dc_comp}e), it suffers from degraded pattern quality and is restricted to use only a subset of subcarriers with compromised CSI structure. These results demonstrate that existing distortion compensation schemes experience significant performance degradation when applied to modern WiFi cards. We also remark that phase-based calibration methods \cite{kotaru2015spotfi, yu2018qgesture,zhu2017phasenoise} address only phase offsets while neglecting magnitude distortions, making them ineffective with modern WiFi hardware.


To address this challenge, we propose Domino, a new framework that enables accurate RF distortion compensation in modern WiFi sensing by transforming CSI into the delay domain and operating on the resulting channel impulse response (CIR). Domino is based on the key insight that hardware-induced distortions affect all signal paths uniformly, allowing the dominant (strongest static) path to serve as a reliable reference for effectively eliminating these distortions. The preliminary efficacy of Domino is demonstrated in Fig.~\ref{fig:dc_comp}f, where the compensated CSI magnitude reveals distinct respiratory cycles that closely align with the ground-truth pattern.  
Our contributions are threefold: (1) we construct a distortion model in the CIR domain that provides theoretical foundations, while also addressing practical constraints of CIR acquisition in COTS systems; (2) we propose a new compensation technique through delay domain processing, enabling selection of a static reference path for effective distortion compensation; and (3) we validate Domino by various real-world respiration monitoring experiments, demonstrating significant improvements in sensing accuracy and robustness compared to conventional approaches.

\section{CIR-based Distortion Model}
\label{cir_model}

In this section, the distortion model in the CIR domain is first introduced. 
Subsequently, we present a method for CIR acquisition from CSI collected with commodity WiFi systems.

\subsection{ Distortion Model }


The undesired phase offset contains multiple contributions \cite{zhu2018pi}. Some of them, including the channel frequency offset (CFO), the phase-locked loop (PPO) and the phase ambiguity (PA), introduce random phase offsets. Meanwhile, sampling frequency offset (SFO) and packet detection delay (PDD) manifest as random delay shifts.  
In addition, AGC induces random gain that fluctuates between measurements. 
The noise-free CIR of the $n$-th tap at a receiving antenna, accounting for hardware distortions, can be expressed as
$$
h[n] = \beta e^{-j\theta}\sum_{l=0}^{L-1} \alpha_l e^{-j 2\pi f_c \tau_l} p[n, \tau_l + \epsilon],
$$
where $\alpha_l$ corresponds to the complex channel gain of the $l$-th path, $\tau_l$ represents the time delay of the $l$-th path, $f_c$ denotes the center frequency, $L$ indicates the total number of multipath components. For hardware-induced distortions, $\beta$ represents the magnitude distortion, $\theta$ represents the phase offset, and $\epsilon$ denotes the delay shift. The function $p[n, \tau_l + \epsilon]$ represents the sampled pulse shaping filter evaluated at the $n$-th tap for the $l$-th path delay ($\tau_l$) shifted by distortion $\epsilon$. This pulse shaping filter implements bandwidth constraints on the transmitted signal, and determines how energy from each path is distributed across nearby CIR taps. 
We can observe that, while these distortions vary randomly between measurements, they affect all multipath components uniformly within each measurement.

\begin{figure}
  \centering
\includegraphics[width=0.85\linewidth]{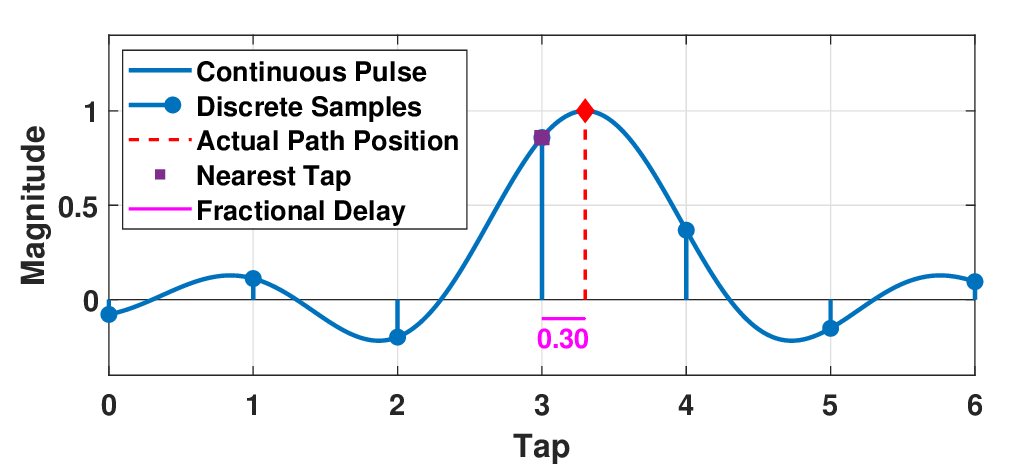}
\caption{Sampling of the pulse with fractional delay}
\vspace{-1em}
\label{pulse}
\end{figure}

In CIR-based signal processing, special attention should be given to the effects of pulse shaping. The pulse-shaping function exerts a considerably pronounced influence on magnitude due to the fractional delay. The fractional delay $\Delta(\tau_l)$ represents the misalignment between the actual propagation delay and the discrete sampling grid. In the presence of hardware distortions, the effective delay becomes $\hat\tau_l = \tau_l + \epsilon$. The fractional delay for the $l$-th path is then defined as: 
$\Delta(\hat\tau_l) = \hat\tau_l - {\rm{round}}(\frac{\hat\tau_l}{T_s})\cdot T_s,$  
where $T_s$ represents the sampling time interval, ${\rm{round}}(\cdot) \mathop: \mathbb{R} \to \mathbb{Z}$ represents the function that rounds each real number to the nearest integer. 
As shown in Fig.~\ref{pulse}, one can easily observe that the magnitude of the nearest tap decreases with the fractional delay increases. It is important to note that the above observations are derived without making any assumptions about the specific pulse shaping filter used, indicating their applicability across various filters.

\subsection{CIR from Commodity WiFi}
\label{cir_measure}

The direct application of CIR-based methods in commodity WiFi systems faces a direct challenge on the acquisition of accurate CIR estimations.  
In practical WiFi systems, CIR is not directly available as an intermediate output and must be derived from CSI measurements, which only contain values of a subset of all subcarriers, so-called active subcarriers. A direct IDFT approach to obtain CIR from these CSI measurements of active subcarriers could lead to inaccurate results. 
Due to the inherent sparsity of wireless channels, the number of taps in the delay domain is often significantly smaller than the DFT length $ N $. Consequently, the least-squares (LS) method can be employed to estimate the CIR from CSI measurements with partial subcarrier usage \cite{kong2024csirff}. 
The LS method estimates the CIR as:
\begin{align}
\label{ls_cirsense}
    \hat{\mathbf{h}} &= (\mathbf{F}_{\mathcal{K},\mathcal{L}}^H \mathbf{F}_{\mathcal{K},\mathcal{L}})^{-1} \mathbf{F}_{\mathcal{K},\mathcal{L}}^H \Tilde{\mathbf{h}}, 
\end{align}
where $\Tilde{\mathbf{h}}$ represents the CSI measurement, $\mathbf{F}$ represents the full unitary DFT matrix, and $\mathbf{F}_{\mathcal{K},\mathcal{L}}$ is the sub-matrix of $\mathbf{F}$, comprising all rows corresponding to the active subcarrier set $\mathcal{K}$ and all columns corresponding to the potential tap set $\mathcal{L}$. The operators $(\cdot)^H$ and $(\cdot)^{-1}$ denote the Hermitian transpose and matrix inverse, respectively. 

\begin{figure}
    \centering
    \includegraphics[width=\linewidth]{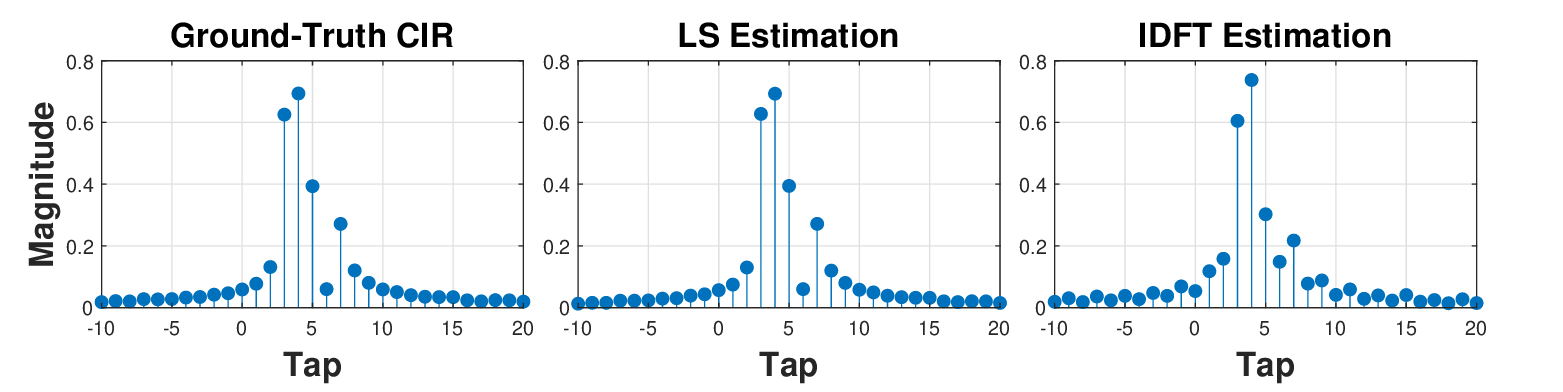}
    \caption{CIR acquisition based on LS and IDFT estimation.}
    \label{fig:cir_est}
    \vspace{-1em}
\end{figure}

A simulation-based evaluation approach is adopted to validate the LS method, given the inherent difficulties in obtaining ground-truth channel information in real-world scenarios. As illustrated in the middle subfigure in Figure~\ref{fig:cir_est}, the LS approach generates tap estimates that closely approximate the ground-truth in the left subfigure. In contrast, the direct IDFT-based estimation results, shown in the right subfigure, demonstrate noticeable deviations from the ground truth.

\section{Dominant Path-based Compensation}
\label{rfdistortion_domino}

To ensure high sensing accuracy and robustness, time-varying random RF distortions must be compensated to ensure accurate motion pattern recognition. 
Our proposed solution to this issue is based on the insight that the magnitude and phase distortions caused by hardware imperfections are consistent across all paths within each CSI or CIR. Therefore, the dominant static path can be used as a reference to align other paths and compensate for RF distortions. Specifically, we propose a two-step solution. 
The first step addresses the random delay $\epsilon$ by aligning the first tap $h[0]$ with the strongest propagation path (dominant path). 
In typical indoor environments, the path exhibiting maximum gain $|\alpha|$ generally corresponds to a static path, characterized by delay $\tau_0$ \cite{meneghello2022sharp}. It typically exhibits higher power compared to other paths. This property, combined with the rapid decay of the pulse shape function $p[n, \tau_l + \epsilon]$ for non-zero differences between $nT_s$ and $\tau_l + \epsilon$, permits the derivation of the following approximation:
\begin{align}
    h[n_0]  &\approx \beta e^{-j\theta}\alpha_0 e^{-j 2\pi f_c \tau_0} p[n_0, \tau_0 + \epsilon] ,
\end{align}
where $n_0$ is the closest tap to $\tau_0+\epsilon$ and $h[n_0]$ is value of the strongest tap. 
The optimal delay shift parameter $\epsilon'_{\text{est}}$ can be determined through the following maximization\footnote{Note that $ \epsilon' $ may take fractional values and the expression $h[n + \epsilon']$ is misused for notation simplicity. The delay-shifting operation can be implemented efficiently in the frequency domain by applying dedicated phase shifts to raw CSI.}:
\begin{align}
\label{losalign}
\epsilon'_{\text{est}} &= \arg\max_{\epsilon'} |h[0 + \epsilon']| \nonumber\\
&\approx \arg\max_{\epsilon'} |\beta e^{-j\theta}\alpha_0 e^{-j 2\pi f_c \tau_0} p[0+\epsilon', \tau_0 + \epsilon]| ,
\end{align}
where the objective function reaches its optimum at $\epsilon' = -\frac{\tau_0+\epsilon}{T_s}$. 
This optimization effectively shifts the CIR to maximize the power concentration in the first tap. After applying this delay shift, the aligned CIR can be expressed~as: 
$$h[n+\epsilon'_{est}] = \beta e^{-j\theta}\sum_{l=0}^{L-1} \alpha_l e^{-j 2\pi f_c \tau_l} p[n, \tau_l - \tau_0].$$
As the second step, the remaining hardware-induced distortions can be eliminated by the following normalization:
\begin{align}
\label{rfcom}
h'[n] &= \frac{h[n+\epsilon'_{est}]}{h[0+\epsilon'_{est}]} \approx \frac{ \cancel{\beta e^{-j\theta}} \sum_{l=0}^{L-1} \alpha_l e^{-j 2\pi f_c \tau_l} p[n, \tau_l - \tau_0]}{ \cancel{\beta e^{-j\theta}} \alpha_0 e^{-j 2\pi f_c \tau_0}} \nonumber\\
&= \sum_{l=0}^{L-1} \alpha'_l e^{-j 2\pi f_c \tau'_l} p[n, \tau'_l] ,
\end{align}
where $h'[n]$ representss the clean version of CIR after distortion mitigation, $\alpha'_l = \frac{\alpha_l}{\alpha_0}$, $\tau'_l = \tau_l-\tau_0$. 
This ratio effectively cancels out the time-varying hardware distortion terms $\beta$ and $\theta$. Since the channel parameters of the dominant static path remain constant over time, any variations in the resulting normalized CIR can be attributed solely to target motion, independent of RF distortions.


   

\section{Systematic Evaluation}
\label{evaluation}

\subsection{Experimental Settings}
\label{sec:exp_setting}

\begin{figure}
  \centering
  \subfloat[LoS setup.]{\includegraphics[width=\linewidth]{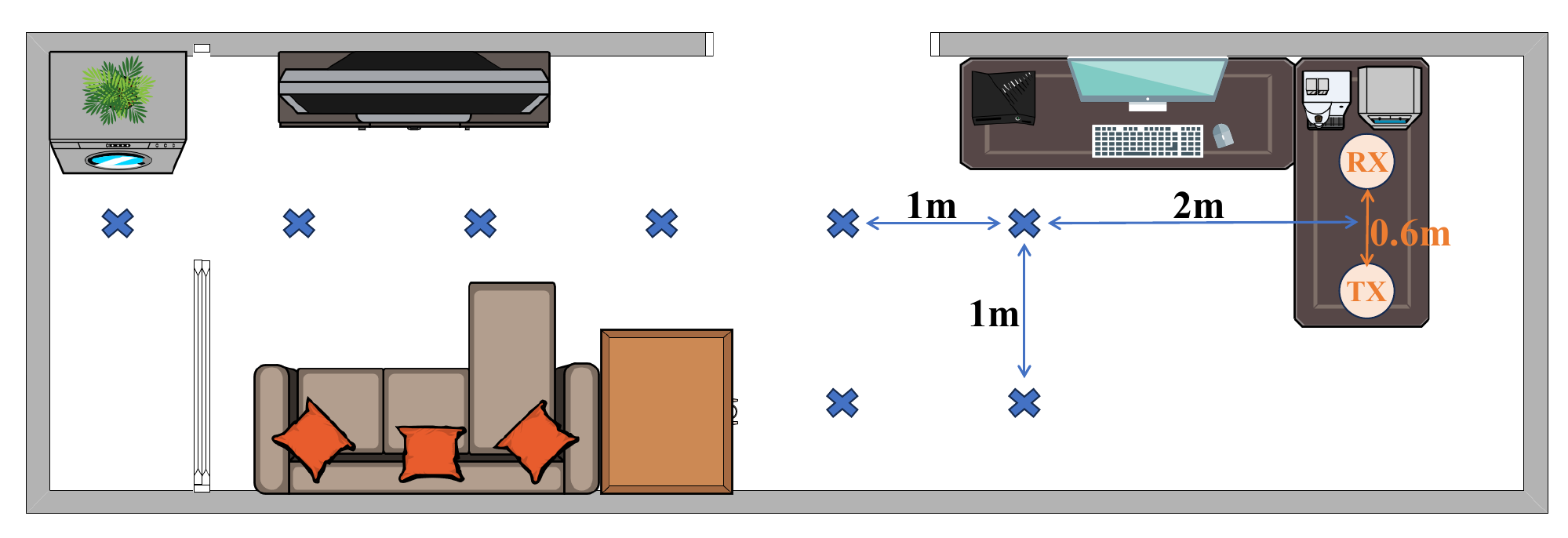}\label{fig:living}}
  
  \subfloat[NLoS setup.]{\includegraphics[width=0.5\linewidth]{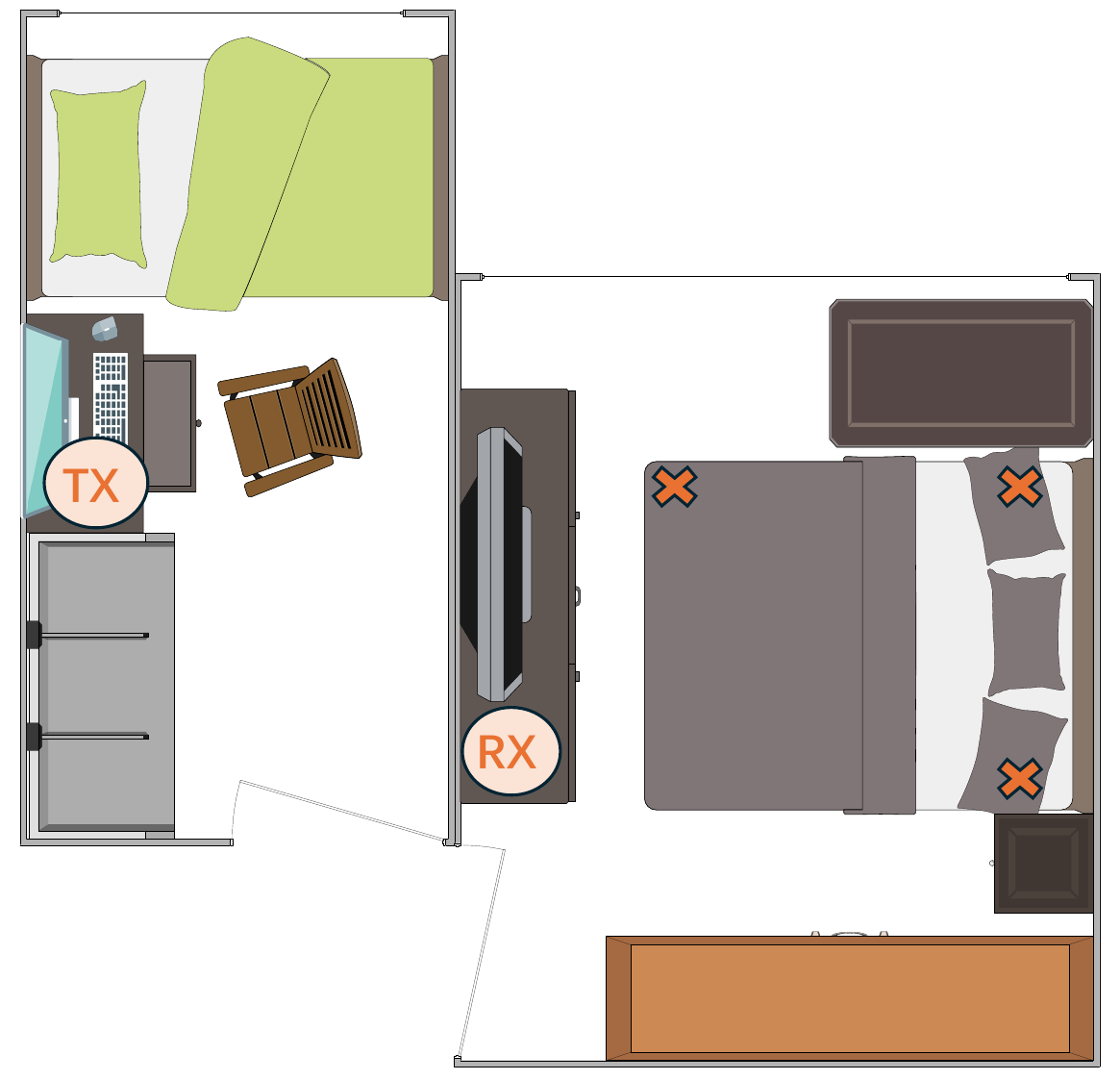}\label{fig:nlos_setup}}
  \subfloat[CDF of errors.]{\includegraphics[width=0.5\linewidth]{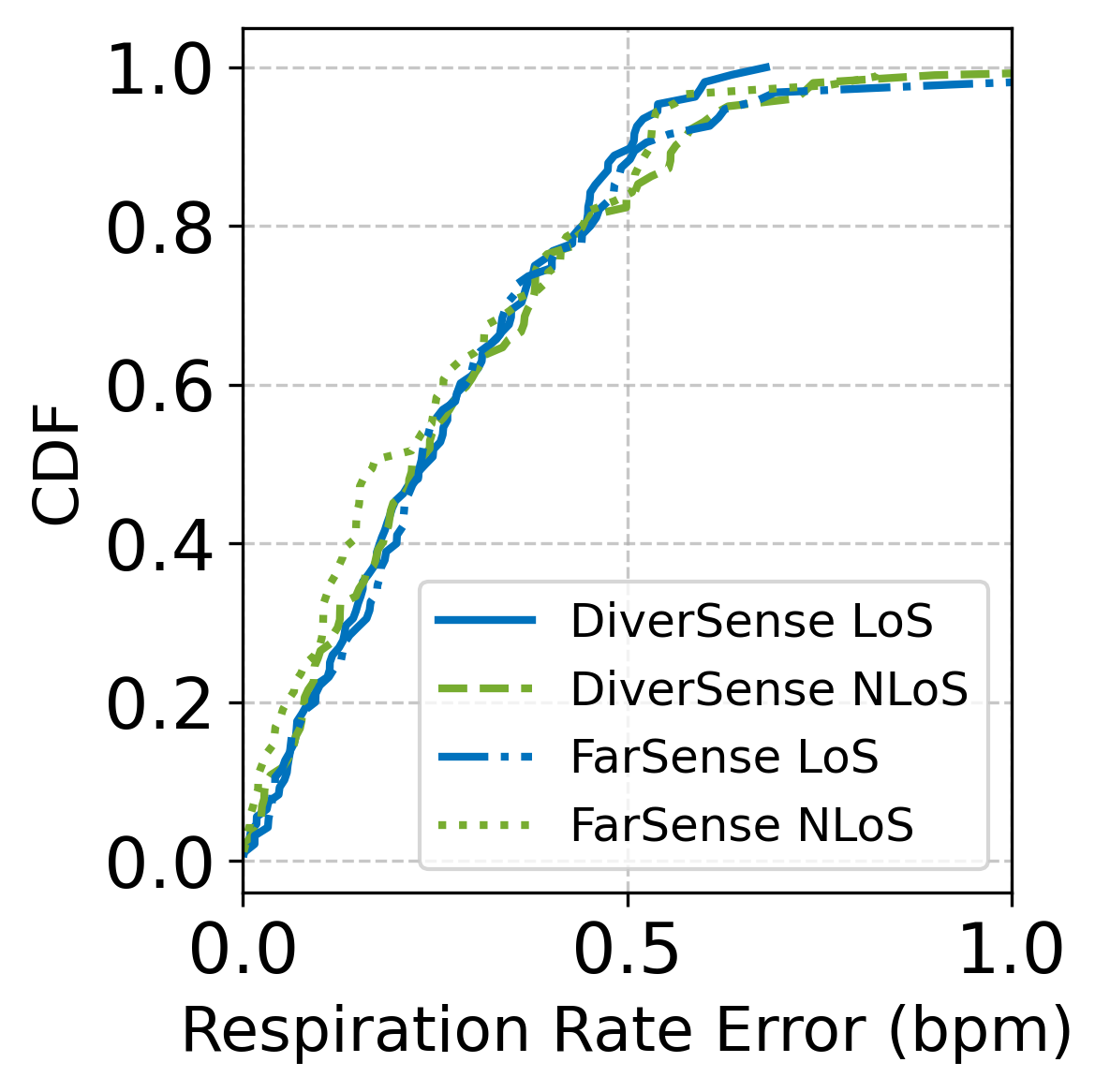}\label{fig:nlos_re}}
\caption{Experimental settings and results.}
\vspace{-1em}
   \end{figure}

To comprehensively evaluate the compensation capability of Domino, we conducted extensive experiments across diverse environmental conditions. Our experimental setup includes transmitter (TX) and receiver (RX) configurations in both line-of-sight (LoS, Fig.~\ref{fig:living}) and non-line-of-sight (NLoS, Fig.~\ref{fig:nlos_setup}) scenarios, where crosses in the figures represent the testing positions of targets. Data collection was repeated multiple times at these testing points across different days to ensure statistical reliability. All CSI measurements were collected using the Picoscenes platform \cite{jiang2021eliminating} deployed on a mini-PC equipped with an Intel AX200 network interface card (NIC) with a single receiving antenna. 
The transmitter is a modified ASUS TUF Gaming AX3000 router that transmit using a single antenna with 200 Hz transmitting rate. The network is configured to operate at 5.25 GHz with a bandwidth of 160 MHz, using the 802.11ax protocol. The ground truth of the respiration rate is collected using a commercial Neulog respiration belt \cite{Belt}. 

\subsection{Sensing Reliability}

To validate the effectiveness of the Domino method in recovering accurate sensing information, we conducted a comprehensive analysis using two state-of-the-art CSI-based respiration monitoring methods: FarSense \cite{zeng2019farsense} and DiverSense \cite{li2022diversense}. We feed these application modules with CSI compensated by Domino and its baselines to evaluate their performance.

The cumulative distribution function (CDF) of respiration rate estimation errors is presented in Fig.~\ref{fig:nlos_re}. 
The key finding from this experiment demonstrates that the two respiration estimation methods achieve high sensing accuracy and maintain consistent performance across both LoS and NLoS scenarios when using Domino-compensated data. Specifically, the median error remains below 0.24 breaths per minute (bpm) for both methods, with 80\% of errors falling below 0.45 bpm, meeting the demanding requirements of practical health monitoring deployments \cite{li2022diversense}. These results confirm that our approach, leveraging the dominant static path identified via CIR analysis, effectively compensates for RF distortions across diverse environments, thereby enabling high sensing accuracy.

\subsection{Comparison of Compensation Schemes}
\label{impact:dis_com}

\begin{figure}
  \centering  
  \subfloat[Evaluation on FarSense \cite{zeng2019farsense}.]{\includegraphics[width=\linewidth]{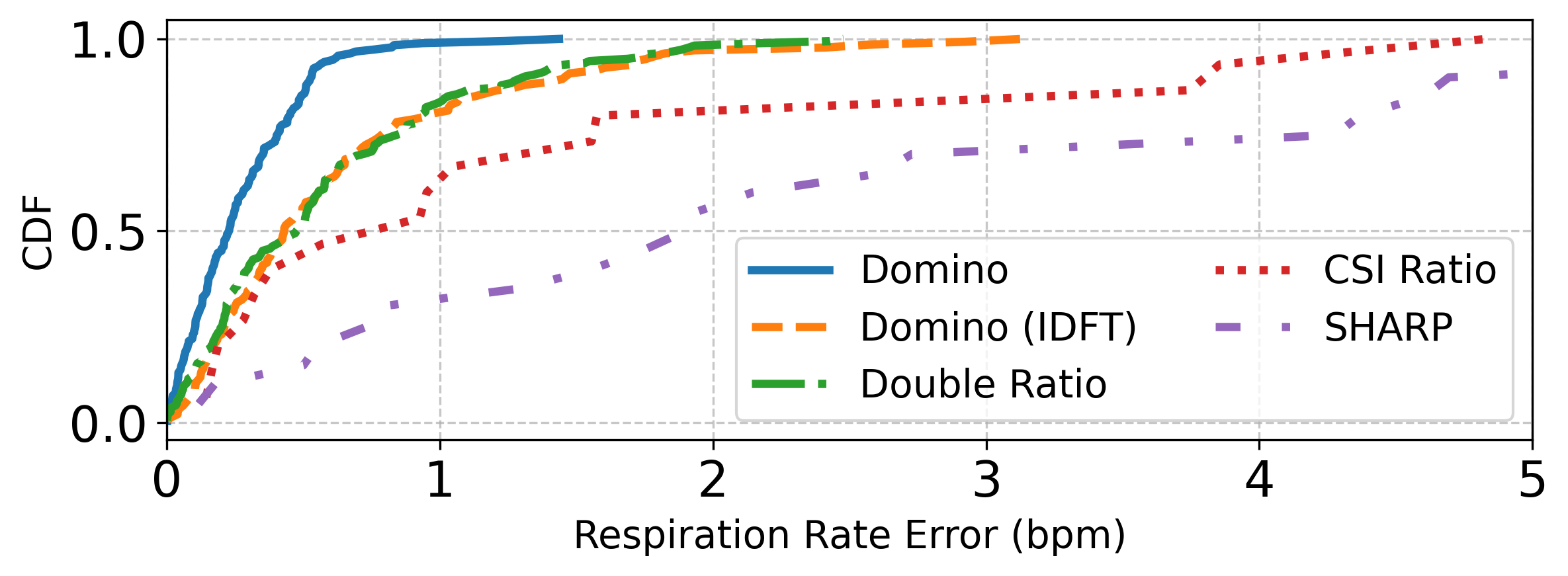}\label{fig:preprocess_f}}
  
  \subfloat[Evaluation on DiverSense \cite{li2022diversense}.]{\includegraphics[width=\linewidth]{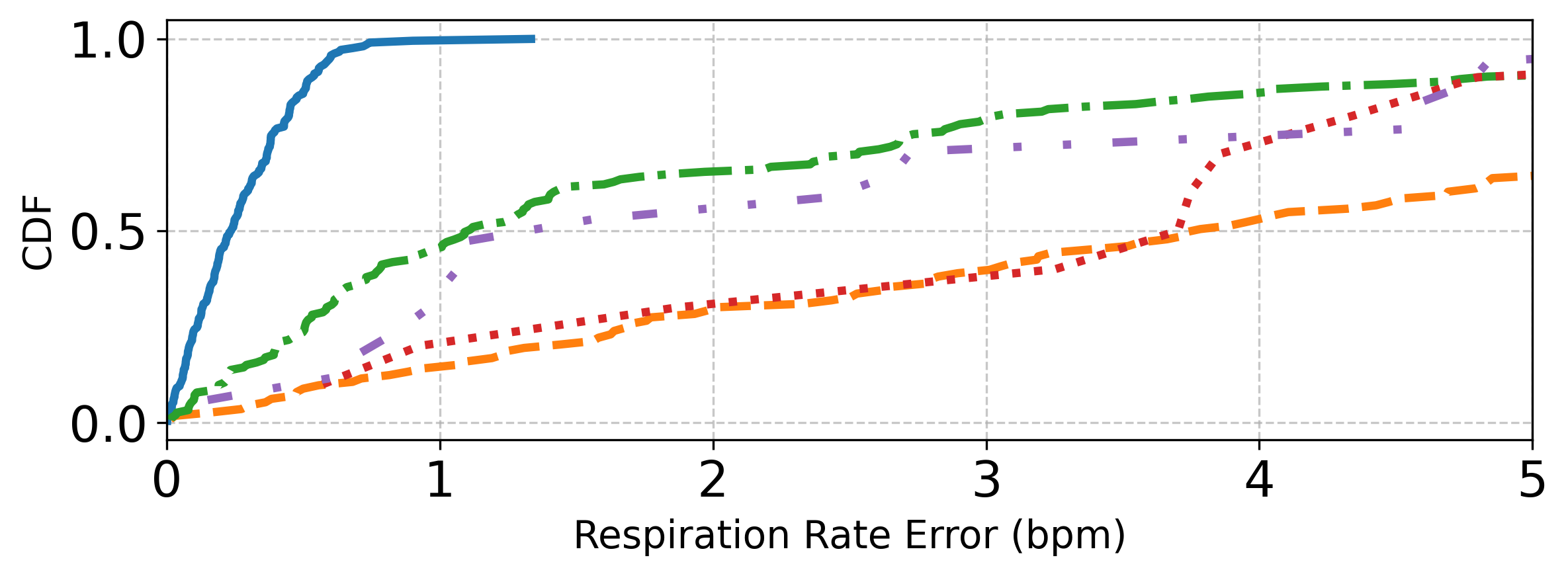}\label{fig:preprocess_d}}
\caption{Comparison of distortion compensation methods.}
\label{fig:compare}
\vspace{-1em}
\end{figure}

Fig.~\ref{fig:compare} presents a comparative analysis of four distortion compensation approaches: Domino (ours), Double Ratio \cite{enabling24yi}, CSI Ratio \cite{zeng2019farsense}, and SHARP \cite{meneghello2022sharp} using all data collected in both LoS and NLoS scenarios. For ablation purposes, we also include Domino (IDFT), which uses the IDFT for CIR estimation rather than the adopted LS approach, to demonstrate the accuracy difference between these estimation methods. 
As illustrated in Fig.~\ref{fig:preprocess_f} and ~\ref{fig:preprocess_d}, Domino consistently maintains superior performance with mean errors remaining below 0.28 bpm across both testing approaches. In contrast, baseline methods show degraded performance. With FarSense, mean errors are: 0.62 bpm for Domino (IDFT), 0.58 bpm for Double Ratio , 1.37 bpm for CSI Ratio, and 2.52 bpm for SHARP. With DiverSense, accuracy further deteriorates: 3.79 bpm for Domino (IDFT), 1.85 bpm for Double Ratio, 3.46 bpm for CSI Ratio, and 2.40 bpm for SHARP. In short, Domino reduces error by at least half compared to baselines.

Additionally, we observe that baselines demonstrate better performance when paired with FarSense compared to DiverSense. This is primarily because DiverSense depends more substantially on preserving pattern accuracy and inter-subcarrier pattern relationships. The baselines inadequately maintain these critical relationships, resulting in notable performance degradation. By comparison, Domino effectively preserves signal integrity throughout the compensation process, demonstrating potential capability beyond respiration monitoring to other sensing tasks, such as target localization.

\vspace{-0.5em}
\section{Conclusions}
\vspace{-0.5em}
This paper presents Domino, a new framework for RF distortion compensation in modern WiFi sensing. By transforming CSI to the delay domain and working with CIR, Domino successfully overcomes limitations of existing compensation methods. Our key insight lies in using the dominant static path as a reference to effectively eliminate distortions through delay domain processing. 
Our real-world experimental evaluation in respiration monitoring demonstrates that Domino significantly outperforms state-of-the-art distortion compensation methods. The approach maintains consistent high accuracy (median error below 0.24 bpm) using a single antenna across both LoS and NLoS environments while preserving the critical signal integrity and inter-subcarrier relationships. 
These results establish Domino as a practical and effective solution for enhancing WiFi sensing reliability in modern systems, addressing the critical transition challenge that has threatened the practical viability of WiFi sensing with new-generation hardware.


\newpage
\bibliographystyle{IEEEbib}
\bibliography{strings,refs}

\end{document}